\documentclass[fleqn,11pt,twoside]{article}

\usepackage{amsthm,amsthm,amssymb, color, xcolor,epsfig, graphics, subfigure}

\usepackage{amsmath, graphicx, latexsym, lscape }

\usepackage{cite}

\makeatletter
\newcommand{\copyrightnote}[2]{{\renewcommand{\thefootnote}{}
 \footnotetext{\small\it
\begin{flushleft}
 \copyright \ #1   #2  
\end{flushleft}}}}

\newcommand{\Name}[1]{\begin{flushleft}
                       \LARGE \bf #1
                       \end{flushleft}\vspace{-3mm}}

\newcommand{\Author}[1]{\begin{flushleft}
                       \it #1 \end{flushleft}}

\newcommand{\Address}[1]{\begin{flushleft}
                       \it #1 \end{flushleft}}

\newcommand{\Date}[1]{\begin{flushleft}
                      \small  \it #1 \end{flushleft}}

\newcommand{\evenhead}{G. Gaeta \& S. Walcher}
\newcommand{\oddhead}{Floquet-like theory and gauge transformations}

\renewcommand{\@evenhead}{
\hspace*{-3pt}\raisebox{-15pt}[\headheight][0pt]{\vbox{\hbox to \textwidth
{\thepage \hfil \evenhead}\vskip4pt \hrule}}}
\renewcommand{\@oddhead}{
\hspace*{-3pt}\raisebox{-15pt}[\headheight][0pt]{\vbox{\hbox to \textwidth
{\oddhead \hfil \thepage}\vskip4pt\hrule}}}
\renewcommand{\@evenfoot}{}
\renewcommand{\@oddfoot}{}

\long\def\@makecaption#1#2{%
  \vskip\abovecaptionskip
  \sbox\@tempboxa{\small \textbf{#1.}\ \ #2}%
  \ifdim \wd\@tempboxa >\hsize
    {\small \textbf{#1.}\ \ #2}\par
  \else
    \global \@minipagefalse
    \hb@xt@\hsize{\hfil\box\@tempboxa\hfil}%
  \fi
  \vskip\belowcaptionskip}

\newcommand{\JNMPnumberwithin}[3][\arabic]{%
  \@ifundefined{c@#2}{\@nocounterr{#2}}{%
    \@ifundefined{c@#3}{\@nocnterr{#3}}{%
      \@addtoreset{#2}{#3}%
      \@xp\xdef\csname the#2\endcsname{%
        \@xp\@nx\csname the#3\endcsname .\@nx#1{#2}}}}%
}

\newcommand{\resetfootnoterule} {
  \renewcommand\footnoterule{%
  \kern-3\p@
  \hrule\@width.4\columnwidth
  \kern2.6\p@}
}

\renewcommand{\footnoterule}{}

\makeatother

\theoremstyle{definition}

\begin{document}

\renewcommand{\evenhead}{ {\LARGE\textcolor{blue!10!black!40!green}{{\sf \ \ \ ]ocnmp[}}}\strut\hfill G Gaeta and S Walcher}
\renewcommand{\oddhead}{ {\LARGE\textcolor{blue!10!black!40!green}{{\sf ]ocnmp[}}}\ \ \ \ \   
Floquet-like theory and smooth dynamical systems}

\thispagestyle{empty}
\newcommand{\FistPageHead}[3]{
\begin{flushleft}
\raisebox{8mm}[0pt][0pt]
{\footnotesize \sf
\parbox{150mm}{{Open Communications in Nonlinear Mathematical Physics}\ \  \ {\LARGE\textcolor{blue!10!black!40!green}{]ocnmp[}}
\ \ Vol.5 (2025) pp
#2\hfill {\sc #3}}}\vspace{-13mm}
\end{flushleft}}

\def\a{\alpha}
\def\b{\beta}
\def\ga{\gamma}
\def\de{\delta}   
\def\eps{\varepsilon}
\def\vphi{\varphi}
\def\la{\lambda}
\def\ka{\kappa}
\def\r{\rho}
\def\s{\sigma}
\def\z{\zeta}
\def\om{\omega}
\def\th{\theta}
\def\vth{\vartheta}
\def\vphi{\varphi}

\def\qbar{\overline{q}}
\def\ombar{\overline{\omega}}
\def\Sbar{\overline{S}}
\def\Hbar{\overline{H}}

\def\A{{\cal A}}
\def\C{{\bf C}}
\def\E{{\cal E}}
\def\F{{\cal F}}
\def\G{{\cal G}}
\def\H{{\bf H}}
\def\Hb{{\bf H}}
\def\h{{\cal H}}
\def\I{{\cal I}}
\def\J{{\cal J}}
\def\Jb{{\bf J}}
\def\K{{\cal K}}
\def\Kb{{\bf K}}
\def\L{\mathcal{L}}
\def\M{{\cal M}}
\def\Mb{{\bf M}}
\def\O{{\cal O}}
\def\P{\mathcal{P}}
\def\Q{{\cal Q}}
\def\R{{\bf R}}
\def\S{{\mathcal S}}
\def\T{{\rm T}}
\def\V{{\cal V}}
\def\W{{\cal W}}
\def\X{{\cal X}}
\def\Z{{\bf Z}}

\def\xb{{\bf x}}
\def\yb{{\bf y}}
\def\zb{{\bf z}}
\def\wb{{\bf w}}

\def\xd{\dot{x}}
\def\yd{\dot{y}}

\def\Ga{\Gamma}
\def\De{\Delta}
\def\La{\Lambda}
\def\Om{\Omega}
\def\Th{\Theta}

\def\pa{\partial}

\def\grad{\nabla}     
\def\lapl{\triangle}  
\def\ss{\subset}
\def\sse{\subseteq}

\def\LRA{\Leftrightarrow}
\def\<{\langle}
\def\>{\rangle}

\def\({\left(}
\def\){\right)}
\def\[{\left[}
\def\]{\right]}
\def\=#1{\bar #1}
\def\~#1{\widetilde #1}
\def\wt#1{\widetilde #1}
\def\.#1{\dot #1}
\def\^#1{\widehat #1}
\def\wh#1{\widehat #1}
\def\"#1{\ddot #1}

\def\bfmu{\boldsymbol{\mu}}

\def\eeq{\end{equation}}
\def\beq{\begin{equation}}

\def\EOR{ \hfill $\odot$}
\def\EOP{ \hfill $\diamondsuit$}

\def\beql#1{\begin{equation} \label{#1}}
\def\eqref#1{(\ref{#1})}

\FistPageHead{1}{\pageref{firstpage}--\pageref{lastpage}}{ \ \ Article}

\strut\hfill

\strut\hfill

\copyrightnote{The author(s). Distributed under a Creative Commons Attribution 4.0 International License}

\Name{Floquet-like theory and gauge transformations for general smooth dynamical systems}

\Author{Giuseppe Gaeta$^{1,2}$ \ and \ Sebastian Walcher$^3$}

\Address{${ }^1$ {\it Dipartimento di Matematica, Universit\`a degli Studi di Milano,  v. Saldini 50, I-20133 Milano (Italy)} and {\it INFN, Sezione di Milano} \\
${ }^2$ {\it SMRI, I-00058 Santa Marinella (Italy)} \\
${ }^3$ {\it Fachguppe Mathematik, RWTH Aachen, D-52056 Aachen (Germany)}}

\Date{Received August 4, 2025; Accepted September 11, 2025}

\setcounter{equation}{0}

\begin{abstract}
\noindent 
The classical Floquet theory allows to map a time-periodic system of linear differential equations into an autonomous one. By looking at it in a geometrical way, we extend the theory to a class of non-autonomous non-periodic equations. This is obtained by considering a change of variables which depends on time in a non-trivial way, i.e. introducing \emph{gauge transformations}, well known in fundamental Physics and Field Theory -- but which seem to have received little attention in this context.
\end{abstract}

\label{firstpage}



\section{Introduction}

It is well known that linear systems of ordinary differential equations in $M=R^n$ with periodic coefficients (with mild non-degeneracy conditions) can be mapped, thanks to \emph{Floquet theory}, into autonomous systems in $M$, identified by a \emph{mo\-no\-dro\-my matrix}.

In this paper we look at Floquet theory in a geometrical fashion, arguing that the fundamental matrix identifies a connection in the full phase space $\wt{M} = M \times S^1$, where $t \in S^1$ due to the periodical nature of the system, seen as a fiber bundle $(\wt{M},\pi, S^1)$ over $S^1$; and that the Floquet map corresponds to a \emph{gauge transformation}. Once the problem is formulated in this way, it is quite natural to consider also cases where the coefficients depend on time in an arbitrary (smooth) fashion; in this case we deal with $\wt{M} = M \times R$ and with a connection in $(\wt{M},\pi,R)$.

That is, in the present paper we discuss how to extend Floquet theory to general, \emph{non-periodic} non-autonomous systems of ODEs.

We would like to stress that in our Examples the relevant gauge transformations are found by inspection -- which is possible thanks to their simple nature. In a companion paper \cite{GW2024} we discuss the solution of the relevant determining equation allowing to determine the gauge transformation for a concrete non-autonomous system (in that paper we also discuss the problem of how our approach deals with nonlinear perturbations of non-autonomous linear systems).

It should be mentioned that there have been, of course, previous works in which gauge transformations were considered in the context of 
finite dimensional dynamics, see in particular \cite{MarSka1}. Our approach is however, as far as we know, quite different from previously considered ones.

\medskip
The \emph{plan of the paper} is as follows. In Sect.\ref{S:FTB} we recall some basic facts of Floquet theory; in Sect.\ref{S:G} we give a geometrical interpretation of it. In Sect.\ref{S:GNP} we argue that the same description can be applied, with appropriate changes, to \emph{non-periodic} systems. We then discuss some simple concrete Examples, starting with (non-uniformly) rotating reference frames in Sect.\ref{S:RRF}; further Examples involve both the Abelian gauge group modelled on SO(2), in Sect.\ref{S:EA}, and the non-Abelian one modelled on SU(2), in Sect.\ref{S:ENA}. Then in Sect.\ref{S:RE} we discuss a (relevant) case in which the gauge group is not given \emph{apriori}, i.e. that of \emph{Riccati equations}. Sect.\ref{S:DC} is then devoted to a brief discussion of relations with the Physics literature; and to draw our conclusions.

\addcontentsline{toc}{subsection}{Acknowledgements}
\subsection*{Acknowledgements}

We are grateful to an anonymous Referee for carefully checking our computations and for spotting several problemas with them. The work of GG is partially supported by the project {\it ``Mathematical Methods in Non-Linear Physics''} (MMNLP) of INFN, and by GNFM-INdAM. In the course of this work GG enjoyed the warm hospitality and the relaxed work atmosphere of SMRI.

\section{Floquet theory}
\label{S:FTB}

In this Section we will briefly recall the main result of Floquet theory.  For a general treatment, see e.g. \cite{JKH,ArnODE,FV}. The theory is often discussed -- for obvious physical reasons -- in the context of second order ODEs, while here we deal with first order ones. Physicists also know this theory under the name of Bloch theory for periodic potentials in Quantum Mechanics; see e.g. \cite{Mermin,Kittel}.

We deal with a \emph{linear} equation
\beql{eq:xd} \xd \ = \ A(t) \ x \ , \ \ \ \ A (t + T ) \ = \ A (t) \eeq
where $x \in R^n$, the dot denotes time derivative, and $A : R_+ \to GL(n;R)$ is a periodic matrix function.

Consider $n$ independent solutions $\xb_{(k)} (t)$ to \eqref{eq:xd}, each of them being a $n$-dimensional vector:
$$ \xb_{(k)} (t) \ = \ \( x_{(k)}^1 (t) , ... , x_{(k)}^n (t) \) \ . $$
The matrix $\Phi (t)$ with entries
$ \Phi^i_k (t) = x_{(k)}^i (t) $ is said to be a \emph{fundamental solution matrix}, or just \emph{fundamental matrix}, for \eqref{eq:xd}.
By construction, the fundamental matrix evolves in time according to the same equation as $x(t)$, i.e.
\beql{eq:Phidot} \dot{\Phi} \ = \ A \ \Phi \ . \eeq
Any solution $\xb (t)$ to \eqref{eq:xd} can be written as
$\xb = \Phi {\bf v}$ with ${\bf v}$ a constant vector, determined by initial conditions.

\medskip\noindent
{\bf Theorem 1.} (Floquet). {\it Each fundamental solution matrix $\Phi (t)$  of eq.\eqref{eq:xd} may be written as
$$ \Phi (t) \ = \ P(t) \ \exp [ B t] \ , $$ where $P(t) : R \to GL(n;R)$ is periodic of period $T$, and $B \in \mathrm{Mat} (n;R)$ is constant.}

\medskip\noindent
{\bf Proof.} See e.g. \cite{ArnODE}, Section 27, or \cite{JKH}.  \EOP
\bigskip

\noindent
With reference to the notation introduced in the statement of Theorem 1, we define
\beql{eq:M} M(t) \ = \ \exp \[ B t \] \ ; \eeq
the constant matrix
$\mathcal{M} = M(T) = \exp [ B T]$ is said to be the \emph{monodromy matrix}. We note that the periodic matrix $P(t)$ appearing in the statement of Theorem 1 can be written as \beql{eq:P} P(t) \ := \ \Phi (t) \ M^{-1} (t) \ . \eeq

\medskip\noindent
{\bf Theorem 2.} (Floquet). {\it Consider equation \eqref{eq:xd}, and let $P$ and $B$ be as in Theorem 1 above. By the change of variables $\xb \to \yb$ defined by
\beql{eq:xtoy} \xb \ = \ P(t) \ \yb \ , \eeq
the time-periodic system \eqref{eq:xd} is mapped into the \emph{autonomous} system
\beq \dot{\yb} \ = \ B \ \yb \ . \eeq}

\medskip\noindent
{\bf Proof.} Theorem 2 follows from a simple computation; this is relevant for our following discussion, and hence we reproduce it here. By \eqref{eq:xd} and \eqref{eq:xtoy}, \beq
\dot{x} \ = \ \dot{P} \, y \ + \ P \, \dot{y} \ = \ A \, x \ = \ A \, P \, y \ . \eeq Rearranging terms, we get the equation for $y(t)$ as
\beql{eq:yd} \yd \ = \ \( P^{-1} \, A \, P \ - \ P^{-1} \, \dot{P} \) \ y \ := \ \wh{A} \ y \ . \eeq

This is completely general, i.e. it holds for whatever choice of the matrix $P (t)$ occurring in the change of variables.\footnote{Hence the use of the notation $\^A$, leaving $B$ for the specific change of variables \eqref{eq:P}.}

However, if we choose exactly the $P(t)$ considered above, see \eqref{eq:P}, we have a more specific form. In fact, in this case we have
$$
P^{-1} \ = \ M  \ \Phi^{-1}  \ , \ \ \
\dot{P} \ = \ \dot{\Phi}  \ M^{-1} \ - \ \Phi  \ M^{-1} \ \dot{M} \ M^{-1} \ ; $$ moreover we note that \eqref{eq:M} implies $[M,B] = 0$ and
$$ \dot{M} \ = \ B \ M \ = \ M \ B \ . $$
Inserting these into \eqref{eq:yd} and recalling \eqref{eq:Phidot}, we obtain the statement:
\begin{eqnarray}
\yd &=& M \, \Phi^{-1} \, \[  A \, \Phi \, M^{-1} \ - \  \( \dot{\Phi} M^{-1} \ - \  \Phi \, M^{-1} \, \dot{M} \, M^{-1} \) \] \ y \nonumber \\
&=& M \, \Phi^{-1} \, \[  A \, \Phi \, M^{-1} \ - \  \( A \, \Phi \, M^{-1} \ - \  \Phi \, M^{-1} \, \dot{M} \, M^{-1} \) \] \ y \nonumber \\
&=& M \, \Phi^{-1} \, \[  \Phi \, M^{-1} \, \dot{M} \, M^{-1} \] \ y \ = \  \dot{M} \, M^{-1} \ y  \ = \  B \ y \ . \end{eqnarray}
This concludes our computation, i.e. the proof for Theorem 2. \EOP

\medskip\noindent
{\bf Remark 1.} We will in general be interested in considering \emph{perturbations} of the linear system \eqref{eq:xd}, i.e. equations of the form
\beql{eq:xdp} \xd \ = \ A(t) \, x \ + \ N (x,t) \ , \eeq
where $N$ is of higher order in $x$ and possibly explicitly dependent on $t$. Proceeding in exactly the same way, we obtain the system
\beql{eq:ydpert} \yd \ = \ B \, y \ + \ F (y,t) \ , \eeq
where $F (y,t)$ is of higher order in $y$ and in general dependent on $t$ and is given by
\beql{eq:F1} F(y,t) \ = \ P^{-1} \, N(P y , t) \ . \eeq
Note that even in the case $\pa_t N (y,t)=0$, an explicit time-dependence will be introduced in $F$ by $P^{-1} = P^{-1} (t)$ and by $P = P(t)$.

We also note, with reference to \eqref{eq:F1} that by construction both $A$ and $P$ (and hence $P^{-1}$) depend on $t$ with a $T$-periodic behavior. Thus, if $N(x,t)$ does not depend explicitly on time, or if it is also $T$-periodic, we end up with a time-independent linear system and a $T$-periodic perturbation. This same remark also shows that if $N$ depends explicitly on time in a \emph{non $T$-periodic way}, then $F$ will also have such a general $t$ dependence.

Finally, note that examples can be built in which $F$ is actually independent of $t$: this is obtained e.g. by choosing $N(x,t) = P(t) N_0 (x)$, see eq.\eqref{eq:F1}. \EOR

\section{A geometrical look at Floquet theory}
\label{S:G}

We have seen that the net effect of the change of variable \eqref{eq:xtoy} considered above has been an affine map between elements of $GL(n;R)$, i.e.
\beq A \ \to \ \^A \ = \ P^{-1} \, A \, P \ - \ P^{-1} \, \dot{P} \ , \eeq
see \eqref{eq:yd}. This corresponds to a \emph{gauge transformation} \cite{Nak,NS,Schutz,CCL,EGH}, and this observation calls for a geometrical understanding of the classical analytical procedure sketched above.\footnote{This seems not to have been discussed in the literature, at least in the framework of differential equations; things are somewhat different when we look at counterparts of Floquet's theory in Physics (Bloch theory), see the brief discussion in Sect.\ref{S:DC}.}

Let us go back at the Floquet construction.
We can look at the fundamental matrix $\Phi (t) = P(t) M(t)$ -- that is, at the set of independent solutions $\xb_{(k)} (t)$ -- as describing a time-varying reference frame; the time evolution in this moving frame is then trivial.

This view, however, in general (that is, unless $B=0$) requires to consider $t \in {\bf R}$, i.e. to lose the essential information that the system is \emph{periodic}, so that we can think $t \in S^1$.

Floquet theorem (see above) suggests to adopt a similar but slightly different approach, i.e. to focus only on the time-periodic part $P(t)$; this defines a time-periodic reference frame. In such a periodic frame the evolution is non-trivial, but the net effect after one period is embodied in the monodromy matrix $\mathcal{M}$, hence -- once we know the period $T$ -- in the simple (i.e. constant) matrix $B$.

Let us discuss this point of view in some more detail.
We know, by Floquet's theorem, that the fundamental matrix $\Phi$, which by definition satisfies the equation \eqref{eq:xd}, is written in the form
\beq \Phi (t) \ = \ P(t) \ M(t) \ = \ P (t) \ \exp[ B t] \eeq
Differentiating, we obtain
\begin{eqnarray*}
\dot{\Phi} &=& \dot{P} \ e^{B t} \ + \ P \ B \ e^{Bt} \\
&=& \dot{P} \ \( P^{-1} \, P \) \ e^{B t} \ + \ P \ B \ \( P^{-1} \, P \) \ e^{B t} \\
&=& \dot{P} \, P^{-1} \ \Phi \ + \ P \, B \, P^{-1} \ \Phi \ . \end{eqnarray*}
On the other hand, we know that, by definition,
$$ \dot{\Phi} \ = \ A \ \Phi \ . $$
Comparing these two expression, we have
\beql{eq:AfromB} A \ = \ \dot{P} \, P^{-1} \ + \ P \, B \, P^{-1} \ . \eeq
We are of course interested in determining the unknown matrix $B$ in terms of the known matrix $A$; this yields
\beql{eq:BfromA} B \ = \ P^{-1} \, A \, P \ - \ P^{-1} \, \dot{P} \ . \eeq

If we interpret $P$ as a connection \cite{Nak,NS,Schutz,CCL,EGH}, this is just the gauge transform of $A$ via $P$. Needless to say, the periodic $P (t)$ is a legitimate connection -- more precisely, $P(t) d t $ is a legitimate connection form -- in the bundle with fiber ${\bf R}^n$ over $S^1$.

\medskip\noindent
{\bf Remark 2.} If we look at \eqref{eq:AfromB} as defining a periodic matrix $A(t)$ in terms of the constant matrix $B$ and of the periodic matrix $P(t)$, this is a dull equation. On the other hand, if we look at \eqref{eq:BfromA} as defining $B$ in terms of the periodic matrices $A(t)$ and $P(t)$, the fact that $B$ is \emph{constant} (and not just periodic) in time is highly nontrivial. This nontrivial fact is of course tantamount to the Floquet theorem. \EOR

\section{Floquet-like theory for non-periodic equations}
\label{S:GNP}

Floquet theory deals with \emph{periodic} equations; here we extend it to the non-periodic case, making use of the point of view discussed in Sect.\ref{S:G} above.

\subsection{Analytical considerations}

We can reverse and somehow trivialize the point of view discussed in the previous Section \ref{S:G}. That is, suppose we start from a \emph{constant coefficients} linear system
\beql{eq:ydotlincc} \dot{\yb} \ = \ B \, \yb \ , \eeq
and we operate a gauge transformation 
\beql{eq:ytox} x \ = \ P \ y \eeq 
(note that here the role of $P$ is not the same as in \eqref{eq:xtoy}, but is reversed) with a periodic matrix function $P(t)$. This will produce a non-autonomous linear system
\beql{eq:xdotlintd} \dot{\xb} \ = \ A(t) \ \xb \ , \eeq
with $A (t)$ the \emph{periodic} matrix function given by \eqref{eq:AfromB}. Actually, our previous discussion (that is, Floquet theorem) shows that \emph{any} time-periodic linear system can be generated in this way from a constant coefficients system.

But, now we can generalize this construction and consider again the system \eqref{eq:ydotlincc} and a gauge transformation \eqref{eq:xtoy}, now with a time-dependent matrix function $P(t)$ having an \emph{arbitrary} time dependence.

This will produce a non-autonomous system \eqref{eq:xdotlintd}, which now is in general \emph{non} periodic. However, by reversing the transformation, it can be mapped back into the constant coefficients system \eqref{eq:ydotlincc}, and hence integrated.

This shows at once that there is a class of time-dependent systems which can be integrated by a suitable gauge transformation. The relevant point is of course if we can recognize that a given system is in this form, hence integrate it. This requires to look at \eqref{eq:AfromB} considering $A$ as a given matrix, and $P(T)$ as an unknown function, as well as $B$ an unknown constant matrix.

Thus we should solve for
\beql{eq:Pd}
\frac{d P}{d t} \ = \ A \, P \ - \ P \, B  \ , \eeq
where both $P$ and $B$ are unknown. In these terms the problem is under-determined, and it will be helpful to consider cases where either one of $P$ and $B$ is given.

\medskip\noindent
{\bf Remark 3.} We note that for any given $B$ and any initial condition for $P$, this system has a unique solution, as follows from the theory of linear differential equations; see \cite{GW2024} for details. \EOR
\bigskip

Actually, the case where $P$ (as well as $A$) is given is trivial: a tentative $B$ is then determined by
$$ B \ = \ P^{-1} \, A \, P \ - \ P^{-1} \, \dot{P} \ , $$ and this $B$ is either time-independent (and our problem is then solved) or time-dependent (in which case we do not have a solution to our problem, but just a transformation to a different equation of the same nature). So the interesting ``intermediate'' case is the one where the target constant matrix $B$ is given and we have to look for the transformation $P(t)$.

Consider next the case where both $A$ and $B$ are given. We then have to solve for $P(t)$ alone, and this amounts to a system of linear equations for its components $P_{ij} (t)$. The usual existence and uniqueness theorem applies; this involves requirements on the matrices $A$ and $B$. Note also that when we want to map from a time-dependent matrix $A$ to a time-independent matrix $B$, equation \eqref{eq:Pd} is non-autonomous.

It should be noted that we can deal in the same way with a \emph{nonlinear} system: if we start from
\beq \dot{\yb} \ = \ B \, \yb \ + \ {\bf F} (\yb) \ , \eeq
say with ${\bf F}$ polynomial, operating the change of variables \eqref{eq:xtoy} we get
\beq \dot{\xb} \ = \ A(t) \, \xb \ + \ {\bf {\mathcal F}} (\xb , t) \ , \eeq
where $A$ is as above, and the nonlinear part is
\beq {\bf {\mathcal F}} (\xb,t) \ = \ P^{-1} \ {\bf F} (P \xb ) \ ; \eeq the time dependence of ${\bf {\mathcal F}}$ originates in the time dependence of $P(t)$.

Note also that if one is able to solve the linear problem, i.e. to determine if $A$ can be obtained from a constant matrix $B$ via \eqref{eq:AfromB}, and has hence determined the corresponding $P(t)$, then it is immediate to check if this same transformation maps the nonlinear terms ${\bf {\mathcal F}} (\xb , t)$ into autonomous nonlinear terms. This latter behavior is of course non-generic; but see next Remark 4.

\medskip\noindent
{\bf Remark 4.} In general, the new nonlinear term ${\bf {\mathcal F}} (\xb,t)$ (see above) has a completely different form from the old ${\bf F} (\yb)$. However, if ${\bf F}$ is covariant under (a representation $T_g$, with $g \in G$, of) a group $G$, i.e.
\beq {\bf F} (T_g \xb , t) \ = \ T_g \, {\bf F} (\xb,t) \ \ \ \ \forall g \in G \eeq and the map $P(t)$ satisfies $P(t) \in T_G$ $\forall t$, then
\beq {\bf \mathcal{F}} (\xb , t) \ = \ {\bf F} (\yb , t) \eeq as follows by a straightforward computation.

In this case, in particular, if ${\bf F}$ is autonomous then also the new nonlinear term ${\bf \mathcal{F}}$ will be autonomous, and our autonomization of the time-dependent linear part will produce an overall autonomous system. \EOR

\subsection{Geometrical considerations}

The geometrical content of our discussion can be summarized as follows.
We are introducing a \emph{covariant derivative} \cite{Nak,NS,Schutz,CCL,EGH}
\beq \nabla_t \ := \ \pa_t \ -  \ P^{-1} \, \dot{P} \ ; \eeq
this also defines a \emph{connection} in $R_+ \times R $ (or $S^1 \times R$ in the periodic case covered by standard Floquet--Bloch theory).

The linear dynamics defined by this covariant derivative is autonomous, in the sense we have
\beq \nabla_t \xb \ = \ B \ \xb \eeq with a constant matrix $B$. In this sense, we have split the problem of solving the equation  $\dot{\yb} = A(t) \yb$ into two problems: first, determine the covariant derivative $\nabla_t$ (i.e. the associated matrix $P$) hence the time-dependent change of frame in which the dynamics becomes autonomous; and then solve for this autonomous dynamics (needless to say, the total difficulty of the problem is conserved, and lies mainly in the first step). See the discussion in Remark 4 above for what this implies for the nonlinear terms and hence the full nonlinear dynamics.

\medskip\noindent
{\bf Remark 5.} The matrix $\mathcal{M} = \exp[B T]$ represents the \emph{holonomy} of the connection $\nabla$ thus defined. It should be stressed that this whole construction is possible -- and nontrivial -- only because the $T$ periodicity of the problem leads to consider bundles over $S^1$ (the time interval $[0,T]$) rather than over $R$ (the time real line). \EOR

\medskip\noindent
{\bf Remark 6.} Setting the problem in a geometrical language allows to appreciate the relations not only with Bloch theory, but also with the \emph{Berry phase} and \emph{Hannay angle}, both for closed paths (which correspond to the periodic case) and for open ones (which correspond to the non-periodic case). See in this respect \cite{Berry,Hannay,Simon,SW,Zak1,Zak2,Zak3,Zak4a,Zak4b,CF,Pati,ZKP,WZ,ASS,FCG,Mos}. \EOR

\section{Examples. Rotating reference frame}
\label{S:RRF}

We start by considering a classical problem in Physics, i.e. the change of reference frame from a fixed (Newtonian) one to a moving one; these will be denoted as $F$ and $M$ respectively. In order to appreciate the role of gauge transformations we need, contrary to what is done in elementary Physics classes, to consider the case where the rotation is not uniform.

The frame $M$ can in general be moving by (time-dependent) translations and rotations w.r.t. $F$; that is, we should consider the Galilei group as the gauge group. To simplify things, we disregard translations and consider the case where the frames share the same origin; thus $M$ is rotating w.r.t. $F$ with (in general, non constant neither in direction nor in modulus) angular velocity ${\bf A}$. With an obvious notation (where derivatives refers to the reference frame they are computed in), given ${\bf v} (t)$ we have
\beql{eq:velMF} \left[ \frac{d {\bf v}}{d t} \right]_F \ = \  \left[ \frac{d {\bf v}}{d t} \right]_M \ + \ {\bf A} \wedge {\bf v} \ . \eeq

\subsection{Example 1}

Considering rotations in full three-dimensional space -- which requires to consider $SO(3)$ as gauge group -- present some computational complications due to the fact $SO(3)$ is not parallelizable and hence we need to introduce local charts (this is not a problem in principles but requires a heavier notation). Computations are usually performed via the Wei-Norman method \cite{WNM1,WNM2,WNM3}, or we can use the embedding of $SO(3)$ in its covering group $SU(2)$ \cite{Gil,Ham}.

In order not to be distracted by such technicalities, consider the restricted two-dimensional problem\footnote{This Example was also considered in \cite{GW2024}.}, i.e. the case where dynamics takes place in the $(x_1,x_2)$ plane (in the $F$ coordinates), and the (by assumption, non-uniform)  rotation is also within this plane, with angular velocity $ \omega  = \omega (t)$. We write
\beq \vartheta (t) := \vartheta(0) + \int_0^t \omega (\tau) \ d \tau \ . \eeq

Passing to the mobile frame we introduce new variables ${\bf w} = (w_1,w_2)$ with
\beq {\bf w} \ = \ R[\vartheta (t)] \ {\bf x} \ , \ \ \ R[\vartheta] \ = \ \begin{pmatrix} \cos \vartheta & - \sin \vartheta \\ \sin \vartheta & \cos \vartheta \end{pmatrix} \ . \eeq Having assumed a non-uniform rotation, we have $\dot{R} \not= 0$.

Let us now consider, to stay within the scope of our discussion, a first order system defined in the fixed frame by \beq \dot{{\bf x}} \ = \ K \ {\bf x} \ + \ \Phi ({\bf x})  \ ; \eeq
here ${\bf x} = (x_1,x_2)$, $K$ is a constant real matrix, and $\Phi$ is the nonlinear part.
In the new variables we have
\begin{eqnarray} \frac{d}{d t} {\bf w} &=& \frac{d}{d t} ( R \, {\bf x} ) \ = \ \dot{R} \, {\bf x} \ + \ R \, \dot{\bf x} \nonumber \\ &=& \dot{R} \, R^{-1} \, {\bf w} \ + \ R \ \( K \, R^{-1} \, {\bf w} \ + \ \Phi (R^{-1} \, {\bf w} ) \) \\
&:=& \ L \, {\bf w} \ + \ \Psi ({\bf w}) \ . \nonumber \end{eqnarray}
Here $L$ is an explicitly time-dependent (as $\dot{R} \not= 0$) real matrix,
\beq L \ = \ \dot{R} \, R^{-1} \ + \ R \,  K \, R^{-1} \ , \eeq
and the nonlinear part $\Psi$ is of course defined by
\beq \Psi ({\bf w}) \ := \ R \, \Phi \( R^{-1} \, {\bf w} \) \ . \eeq
Note that if $\Phi$ is covariant, $\Phi (R \xb) = R \Phi (\xb)$, then $\Psi ({\bf w}) = \Phi ({\bf w})$.

\section{Examples. The Abelian case -- SO(2)}
\label{S:EA}

In this Section we consider some simple examples in ${\bf R}^2$ with the abelian gauge group $SO(2)$; these are of course related to the discussion given in the previous Sect.\ref{S:RRF}.

Note however the difference with Example 1: in that case we considered a \emph{given} mobile reference frame and considered how the system -- autonomous in the ''original'' one -- looks in the mobile frame; here we will be  looking for a gauge transformation taking the system (non-autonomous in the frame in which it is given) to an autonomous form.

\subsection{Example 2}

Let $(\xi,\eta)$ be cartesian coordinates in ${\bf R}^2$, and consider first the equations
\beql{eq:ex1full} \begin{cases}
\xi' \ = \ \om (t) \, \eta \ + \ (1 \, - \, \xi^2 \, - \, \eta^2 ) \, \xi \ , &\\
\eta' \ = \ - \om (t) \, \xi \ + \ (1 \, - \, \xi^2 \, - \, \eta^2 ) \, \eta \ . & \end{cases} \eeq
This simple system describes a linear rotation with time-dependent angular velocity $\om (t)$ and a nonlinear radial motion driving the system towards the circle of radius $r=1$. Here $\om (t)$ is an arbitrary smooth function of $t$, so the system is non-autonomous.\footnote{The system would of course simplify passing to polar coordinates, but it would then retain its non-autonomous character.}


According to our scheme, we should first isolate the linear part of the system
\beql{eq:ex1lin} \begin{cases}
\xi' \ = \ \xi \ + \ \om (t) \, \eta \ , &\\
\eta' \ = \ - \om (t) \, \xi \ + \ \eta \ ; & \end{cases} \eeq
and now look for a gauge transform taking this to an autonomous system.

We write $\Xi = (\xi,\eta)$ and \eqref{eq:ex1lin} in the form
\beql{eq:ex1Xi} \Xi' \ = \ A (t) \, \Xi \eeq
with
\beq A \ = \ \begin{pmatrix} 1 & \om(t) \\ - \om (t) & 1 \end{pmatrix} \ . \eeq
Note that $\Xi$ can be determined explicitly here, up to a quadrature.

Now we should look for a time-dependent regular linear map
\beql{eq:ex1map} \Xi \ = \ P (t) \ X \ , \eeq where $X = (x,y)$. Under this map, \eqref{eq:ex1Xi} will be transformed into
\beq X' \ = \ \( P^{-1} \, A \, P \ - \ P^{-1} \, P' \) \ X \ := \ B \ X \ , \eeq
as follows from our general discussion. We should look for a $P= P(t)$ such that $B$ is a constant matrix. By definition the matrix $P$ satisfies eq. \eqref{eq:Pd}, that is $ P' = A P - P B$.

We stress that here $A$ is assigned, while $B$ can be \emph{any} constant matrix; it represents, in a way, the target dynamics. In some cases, physical or group-theoretical considerations can lead one to choose a tentative target dynamics, thus simplifying the problem.

In the present case, we may try to eliminate the time-dependence modifying only the $\om (t)$ term; this means we may e.g. aim at
\beql{eq:ex1B} B \ = \ \begin{pmatrix} 1 & 0 \\ 0 & 1 \end{pmatrix}. \eeq
This not only represents a ``minimal'' modification (of the linear part), but also -- not affecting the radial part of the dynamics -- can in principle be reached by an orthogonal transformation, i.e. one of the form
\beql{eq:ex1La} P (t) \ = \ \begin{pmatrix} \cos \b(t) & - \sin \b (t) \\ \sin \b (t) & \cos \b (t) \end{pmatrix} \eeq where $\b (t)$ is a smooth function to be determined.

By \eqref{eq:ex1map}, with $P$ of the form \eqref{eq:ex1La} and $B$ as prescribed by \eqref{eq:ex1B}, the equation for $P$ reads as the system\footnote{The matrix $P$ has four elements and hence the matrix equation for $P$ corresponds to a four dimensional system; however the special form of our initial equation \eqref{eq:ex1full} -- note that $A(t)$ and $B$ commute -- and our \emph{ansatz} \eqref{eq:ex1La} for $P$ make that we have two pairs of identical equations.}
\beq \begin{cases}
 \sin [\beta (t)] \ \beta '(t) \ = \ - \, \omega (t) \ \sin [\beta (t)] \ , & \\
 \cos [\beta (t)] \ \beta '(t) \ = \ - \,  \omega (t) \ \cos [\beta (t)] \ . &  \end{cases} \eeq
This reduces to $\b' (t) = - \, \om (t)$, i.e. is solved by
\beql{eq:ex1solLa1} \b (t) \ = \ \b_0 \ - \ \int_0^t \om (\tau) \ d \tau \ . \eeq


We have thus solved our problem for \eqref{eq:ex1full} at the linear level. We should now go back to consider the full problem,  hence we have to act on our system by the map \eqref{eq:ex1solLa1}.

In this case, we obtain an autonomous nonlinear system
\beq \begin{cases} 
x' \ = \ \( 1 \ - \ (x^2 + y^2) \) \ x \ , & \\
y' \ = \ \( 1 \ - \ (x^2 + y^2) \) \ y \ . & \end{cases} \eeq

Note that the nonlinear term has not changed at all. Obviously, this is related to the fact the nonlinear terms had a very specific autonomous form, i.e. is covariant under orthogonal transformations, and we have indeed been considering an orthogonal transformation $P(t) \in O(2)$ $\forall t$.

\subsection{Example 3}

We will consider a slight variation on Example 2, i.e. a system of the same general form but in which the nonlinear term also has a time dependence:
\beql{eq:ex2full} \begin{cases}
\xi' \ = \ \om (t) \, \eta \ + \ [1 \, - \, R^{-1}(t) (\xi^2 \, - \, \eta^2 ) ] \, \xi \ , &\\
\eta' \ = \ - \om (t) \, \xi \ + \ [1 \, - \, R^{-1}(t) (\xi^2 \, - \, \eta^2 ) ] \, \eta \ . & \end{cases} \eeq
This simple system describes a linear rotation with time-dependent angular velocity $\om (t)$ and a nonlinear radial motion driving the system towards the circle of radius $r=R(t)$. Here $\om (t)$ and $R(t) > 0$ are arbitrary smooth functions of $t$.

Note that again the nonlinear term
${\bf F} \( \Xi , t \) = - (\rho^2/R) \Xi $ is covariant under orthogonal transformations.

The linear part of the system is exactly as in Example 1, so we can proceed in the same way and make this autonomous. We will just consider the case where the linear part is taken to the form $B_1$ (the case $B_2$ is completely analogous).
By the map \eqref{eq:ex1map}, \eqref{eq:ex1La}, \eqref{eq:ex1solLa1}, the system \eqref{eq:ex2full} is transformed into
\beq \begin{cases} x' \ = \ x \ - \ R^{-1} (t) \ \( x^2 \, + \, y^2 \) \ x \ , & \\
 y' \ = \ y \ - \ R^{-1} (t) \ \( x^2 \, + \, y^2 \) \ y \ . & \end{cases} \eeq
Thus, as expected (see the previous considerations on covariant nonlinear terms), we have mapped the system into a system with autonomous linear part and a non-autonomous nonlinear part which has the same form as in the original system \eqref{eq:ex2full}. Note this non-autonomous system has a specially simple form, in that it describes purely radial motions; that is, in polar coordinates we have $$ \theta' \ = \ 0 \ , \ \ \ \rho' \ = \ \( 1 \ - \ \frac{\rho^2}{R(t)} \) \ \rho \ . $$
We stress (just to avoid any possible confusion) that this refer to the ``new'' $(x,y)$ coordinates, i.e. $\rho = \sqrt{x^2+y^2}$, $\theta = \arctan (y/x)$; and should not be confused with the previously introduced ones, $(\vartheta,\varrho)$ referring to the ``original'' $(\xi,\eta)$ coordinates.

\subsection{Example 4}

Finally, we consider a case where the nonlinear part is autonomous and does \emph{not} have a covariant form; we choose e.g.
\beql{eq:ex3full} \begin{cases}
\xi' \ = \ \om (t) \, \eta \ + \ [1 \, - \, \xi \, \eta ] \, \xi \ , &\\
\eta' \ = \ - \om (t) \, \xi \ + \ [1 \, - \, \xi \, \eta ] \, \eta \ . & \end{cases} \eeq

The linear part is still the same as in Examples 1 and 2, and so we operate with the same map \eqref{eq:ex1map}, \eqref{eq:ex1La}, \eqref{eq:ex1solLa1}. In this case its effect on the full system is to map \eqref{eq:ex3full} into
\beq \begin{cases}
x' \ = \ x \ + \ \left[(1/2) \, \sin (2 \beta (t))  \,  \left(y^2-x^2\right) \ - \ \cos (2 \beta (t)) \, x \, y \right] \ x \ , & \\
y' \ = \ y \ + \   \left[(1/2) \, \sin (2 \beta (t)) \,
   \left(y^2-x^2\right)  \ - \ \cos (2 \beta (t)) \, x \, y \right] \ y \ . & \end{cases} \eeq
In this case, not only the nonlinear part has changed under our gauge  transformation, but the system -- which in its original form \eqref{eq:ex3full} had a non-autonomous linear part and an autonomous nonlinear one -- does now have an autonomous linear part and a non-autonomous nonlinear one.

\subsection{Example 5}
\label{sec:NPO}

The previous Examples are in a way abstract ones; we want now to discuss a specific one, i.e. an oscillator with a time dependent frequency; more specifically, we will consider the case where this frequency is quasi-periodic in time. The system we will consider can be integrated in an elementary way, so that the correctness of the result obtained by appying our approach can be checked directly.

Let $\Omega (t)$ denote the function
\beq \Omega (t) \ = \ \om \ + \ a \, \sin (\a t) \ + \ b \, \sin ( \b t ) \ , \eeq
where $a,b$ as well as $\om, \a , \b$ are nonzero real constants, and we will focus on the case where $\a/\b$ is  irrational; thus $\Omega (t)$ is a quasi-periodic function. 

Consider the dynamical system
\beql{eq:exa0}
\begin{cases} \xd \ = \ - \, \Omega (t) \, y \ , & \\
\yd \ = \ \Omega (t) \, x \ . & \end{cases} \eeq
If $a$ and $b$ are small this represents a (quasi-periodic) perturbation of a harmonic oscillator with frequency $\omega$. 

The system is promptly integrated by passing to polar coordinates $(\rho, \theta)$; in these we have $\rho = const$ and 
\beql{eq:exasol} \theta (t) \ = \ \theta (0) \ + \ \int_0^t \Omega (\tau) \, d \tau \ . \eeq
We want however, as mentioned above, to use this simple example to illustrate the application of our method in a concrete and physically meaningful case.

Our system \eqref{eq:exa0} is written in vector notation as 
\beq \dot{\xb} \ = \ A(t) \ \xb \ , \eeq
where $\xb = (x,y)$ and 
\beq A(t) \ = \ \Omega (t) \ J \ , \ \ \ J \ = \ \begin{pmatrix}0 & -1 \\ 1 & 0 \end{pmatrix} \ . \eeq

In order to handle the system with our method, we have to consider eq.\eqref{eq:Pd}, 
where both $P (t)$ and the constant matrix $B$ are unknown. We will make an \emph{ansatz} for $B$ (see also Remark 3 in this respect), i.e. that it corresponds to the unperturbed problem obtained for $a=b=0$,
$$ B \ = \ \omega \ J \ . $$ 
With this choice, our gauge-determining equation reduces to 
\beql{eq:exagde} \frac{d P}{d t} \ = \ \Omega (t) \ J \, P \ - \ \omega \ P \, J \ . \eeq

We will thus look for a solution in the form 
\beql{eq:exaP} P(t) \ = \ \begin{pmatrix} \cos [\phi (t)] & - \sin [\phi(t)] \\ \sin[\phi(t)] & \cos[\phi (t)] \end{pmatrix} \ . \eeq
With $P(t)$ of this form, the equation \eqref{eq:exagde} reduces to 
\beq - \ \( \phi' (t) \ - \ \Omega (t) \ + \ \omega \) \ \begin{pmatrix} \sin [\phi (t)] &  \cos [\phi (t)] \\
- \, \cos [\phi (t)] & \sin [\phi (t)] \end{pmatrix} \ ; \eeq
that is, $\phi (t)$ is determined by solving
\beq \phi' (t) \ = \ \Omega (t) \ - \ \omega \ = \ A \, \sin ( \a t) \ + \ B \, \sin (\b t) \ . \eeq
In conclusion, setting to zero the inessential integration constant, we get 
\beq \phi (t) \ = \ - \, \frac{A}{\a} \, \cos (\a t) \ - \ \frac{B}{\b} \, \cos (\b t) \ ; \eeq
plugging this into \eqref{eq:exaP} we obtain the required gauge transformation. In the new variables \beql{eq:exaCOV}  \begin{pmatrix} \xi \\ \eta \end{pmatrix} \ = \  P \ \begin{pmatrix} x \\ y \end{pmatrix} \eeq the system reads just
\beq \begin{cases} \dot\xi \ = \ - \omega \, \eta \ , & \\ \dot\eta \ = \ \omega \, \xi \ . & \end{cases} \eeq
We can them write the solution for these variables, and go back to the original ones by inverting \eqref{eq:exaCOV}; we obtain of course \eqref{eq:exasol}.

The computation is easily generalized to the case where the nonautonomous term is given by the superposition of $n$ irrational frequencies.

\section{Examples. The non-Abelian case -- SU(2)}
\label{S:ENA}

In this Section we consider simple examples in ${\bf R}^4$ with the non-abelian gauge group $SU2)$. Note we are dealing with real vector fields, so we consider the basic real representation of $SU(2)$ -- or more precisely one of its two fundamental representations, differing for orientation \cite{Kirillov,GRHH}. This is fully equivalent to the standard complex representation in terms of Pauli matrices.

The generators for such a representation -- more precisely for the representation of the Lie algebra $su(2)$ -- can be chosen to be
\begin{eqnarray*}
Y_1 &=& \left( \begin{matrix} 0 & 1 & 0 & 0 \\ -1 & 0 & 0 & 0 \\ 0 & 0 & 0 & 1 \\ 0 & 0 & -1 & 0 \end{matrix} \right) \ , \ \
Y_2 \ = \ \left( \begin{matrix} 0 & 0 & 0 & 1 \\ 0 & 0 & 1 & 0 \\ 0 & -1 & 0 & 0 \\ -1 & 0 & 0 & 0 \end{matrix} \right) \ , \\
Y_3 &=& \left( \begin{matrix} 0 & 0 & 1 & 0 \\ 0 & 0 & 0 & -1 \\ -1 & 0 & 0 & 0 \\ 0 & 1 & 0 & 0  \end{matrix} \right) \ . \end{eqnarray*}
These satisfy
$$ Y_i^2 \ = \ - I \ , \ \ Y_i \, Y_j \ = \ \epsilon_{ijk} \, Y_k \ , \ \ [ Y_i , Y_j ] \ = \ 2 \, \epsilon_{ijk} \, Y_k \ ; $$ here $I$ is the four-dimensional identity matrix, $[.,.]$ is the commutator and $\epsilon_{ijk}$ is the completely alternating (Levi-Civita) symbol.

Given a generic matrix
$$ A \ = \ \sum_{i=1}^3 \a_i \, Y_i \ , \ \ \ \a_1^2 + \a_2^2 + \a_3^2 = 1 \ , $$ such that the we get immediately
$$ A^{2 k} \ = \ (-1)^k \, I \ , \ \ \ A^{2k+1} \ = \ (-1)^k \, A \ . $$
This also guarantees that
$$ \exp[\om A t] \, \xb \ = \ \( \cos (\om t) \, I \ + \ \sin (\om t) \, A \) \ \xb \ . $$
Any element $g \in SU(2)$ can be represented in this way; on the other hand, if we want to consider the \emph{gauge} action of $SU(2)$, we need to promote the constants $\a_i$ to be smooth functions of $t$.

As for the corresponding representation of the group $SU(2)$, this is made of matrices in the form
$$ M \ = \ \left(
\begin{array}{llll}
 a_{0} & a_{1} & a_{3} &  a_{2} \\
 -a_{1} & a_{0} & a_{2} &  -a_{3} \\
 -a_{3} & -a_{2} & a_{0} & a_{1} \\
 -a_{2} & a_{3} & -a_{1} & a_{0}
\end{array}
\right) \ , \ \ \ \sum_{\mu=0}^3 a_\mu^2 \ = \ 1 \ ; $$
writing $Y_0 = I$, these are conveniently written as
$$ M \ = \ \sum_{\mu=0}^3 a_\mu \ Y_\mu \ . $$

Let us now consider the linear, constant coefficients, system
\beql{eq:NA1} \xd^i \ = \ L^i_{\ j} \, x^j \ . \eeq
Under a change of variables $x^i = M^i_{\ j} (t) y^j$, as discussed in Section \ref{S:G}, this is mapped into
\beql{eq:NA2} \yd^i \ = \ \( M^{-1} \, L \, M \ - \ M^{-1} \, \dot{M} \)^i_{\ j} \ y^j \ := \ \wt{M}^i_{\ j} \, y^j \ . \eeq
Conversely, a system of this form is brought to the form \eqref{eq:NA1} by such a transformation.

\subsection{Example 6}

As a concrete example, let us consider the system
\beql{eq:NA3} \yd \ = \ K (t) \ y \eeq
where $K=K(t)$ is given by the matrix
\beql{eq:NA4} K \ = \ \frac{\eta}{3 - \cos^2 (t)} \ \left(
\begin{array}{llll}
 0 & -1 & \cos (\eta  t) & \sin (\eta  t)
   \\
 1 & 0 & \sin (\eta  t) & -\cos (\eta  t)
   \\
 -\cos (\eta  t) & -\sin (\eta  t) & 0 & -1
   \\
 -\sin (\eta  t) & \cos (\eta  t) & 1 & 0
\end{array}
\right) \eeq
with $\eta$ a real parameter. Note that for $\eta$ irrational, this is not periodic. We will also write, for ease of notation, $\om := 1 + \eta$; if $\eta$ is rational (irrational), so is $\om$.

This is mapped into the simple system
\beql{eq:NA5} \xd \ = \ L \ x \ , \ \ \ L \ = \ \left(
\begin{array}{llll}
 0 & -1 & 0 & 0 \\
 1 & 0 & 0 & 0 \\
 0 & 0 & 0 & -1 \\
 0 & 0 & 1 & 0
\end{array}
\right) \eeq using the gauge transformation
$$ x \ = \ M(t) \ y $$ with $M$ as above and
$$ \a_0 (t) \ = \ \frac{\sin (t)}{\sqrt{2}} \ , \ \ \a_1 (t) \ = \ \frac{\cos(t)}{\sqrt{2}} \ , \ \ \a_2 (t) \ = \ \frac{\sin (\om t)}{\sqrt{2}} \ , \ \
\a_3 (t) \ = \ \frac{\cos (\om t)}{\sqrt{2}} \ . $$
In other words, we have
\beql{eq:NA6} M \ = \ \frac{1}{\sqrt{2}} \ \left(
\begin{array}{llll}
 \sin (t) & \cos (t) & \cos (\om t) &
   \sin (\om t) \\
 -\cos (t) & \sin (t) & \sin (\om t)
   & -\cos (\om t) \\
 -\cos (\om t) & -\sin (\om t)
   & \sin (t) & \cos (t) \\
 -\sin (\om t) & \cos (\om t) &
   -\cos (t) & \sin (t)
\end{array}
\right) \ . \eeq

\subsection{Example 7}

We consider a related example; now we write
$$ a \ := \ \eta + 2 \ , \ \ \ b := \eta - 2 \ , $$ with $\eta \not= 0$ again a real constant, and deal with \eqref{eq:NA3} where 
\beql{eq:NA7} K \ = \ \frac{1}{3 - \cos^2 t} \ \left(
\begin{array}{llll}
 0 & - a & a \cos (\eta  t) &
   a  \sin (\eta  t) \\
 a & 0 & a \sin (\eta  t) &
   - a \cos (\eta  t) \\
 - a \cos (\eta  t) & - a
   \sin (\eta  t) & 0 & - b  \\
 - a \sin (\eta  t) & a \cos
   (\eta  t) & b & 0
\end{array}
\right) \ . \eeq 
In this case acting with the same transformation -- i.e. with the same coefficients $\a_i (t)$ as above -- the system is brought to the form
\beql{eq:NA8} \xd \ = \ L \ x \ , \ \ \ L \ = \  \left(
\begin{array}{llll}
 0 & -1 & 0 & 0 \\
 1 & 0 & 0 & 0 \\
 0 & 0 & 0 & 1 \\
 0 & 0 & -1 & 0
\end{array}
\right) \ . \eeq
Note this satisfies
\beql{eq:NA9} \[ Y_i , L \] \ = \ 0 \ , \ \ \ i=1,2,3 \ ; \eeq
thus the effect of the transformation is only in the affine term $M^{-1} \dot{M}$, see \eqref{eq:BfromA}.

\section{Examples. Riccati equations}
\label{S:RE}

The Examples seen above used a \emph{given} gauge group\footnote{As we consider \emph{linear} gauge transformations alone, this actually means a given subgroup of $GL(n;R)$, such as $SO(2) \ss GL(2;R)$ or $SU(2) \ss SO(4) \simeq GL(4;R)$.}. But, in general, when we are faced with a given non-autonomous equation or system we do not know which gauge group (that is, which subgroup of $GL(n;R)$, as we are considering gauged \emph{linear} transformations) could be used to autonomize it.

Here we will consider a relevant class of non-autonomous quadratic equations, i.e. \emph{Riccati equations}, which can be cast as a system of two first order linear equations, and discuss how our approach applies to them.

An equation of the form
\beql{eq:Riccati} \frac{dy}{dt}  \ = \ f(t) \ + \ g(t) \, y \ + \ h (t) \, y^2 \eeq is said to be a Riccati equation. It is convenient, to avoid degeneration, to assume $h(t)$ to be nowhere zero.

\subsection{Mapping to a linear system}

As well known, a Riccati equation \eqref{eq:Riccati} can be mapped to a two-dimensional linear system
\beql{eq:RL}
\frac{d}{dt}\begin{pmatrix} u \\ v \end{pmatrix} \ = \ \begin{pmatrix} g(t) + \a (t) & f(t) \\ - h(t) & \a (t)  \end{pmatrix} \ \begin{pmatrix} u \\ v \end{pmatrix} \ , \eeq where $\a (t)$ is an arbitrary smooth function.
In fact, it is immediate to check that if $\{ u(t),v(t)\}$ is a solution to \eqref{eq:RL}, then $$ y (t) \ = \ \frac{u(t)}{v(t)} $$ is a solution to \eqref{eq:Riccati}.

We can then deal with linear equations of the form \eqref{eq:RL}, identified -- choosing $\a (t) = 0$ for simplicity --  by the matrix
\beql{eq:RA} A(t) \ = \ \begin{pmatrix} g(t) & f(t) \\ - h(t) & 0 \end{pmatrix} \ , \eeq and consider gauge transformations generated by $P(t) : R \to GL(2,R)$.

\medskip\noindent
{\bf Remark 7.} We note in passing that a similar property holds for matrix Riccati equations, which we write as
\beql{eq:RicMat} \dot{Y} \ = \ - \, Y \, M_{21} \, Y \ + \ (M_{11} \, Y \ - \ Y \, M_{22} ) \ + \ M_{12}  \eeq with $Y$ and $M_{ij}$ square $N$-dimensional matrices.

In fact, the rational map
$X\mapsto X_1 X_2^{-1}$ (with $X_i$ square $N$-matrices) sends solutions of the linear (matrix) equation
\beq
\dot X \ = \ \begin{pmatrix} \dot{X}_1 \\ \dot{X}_2 \end{pmatrix} \ = \ \begin{pmatrix} M_{11} & M_{12}  \\
M_{21} & M_{22} \end{pmatrix} \ \begin{pmatrix} X_1 \\ X_2 \end{pmatrix} \eeq
to solutions of the (matrix) Riccati equation \eqref{eq:RicMat}, see \cite{Reid}. \EOR

\medskip\noindent
{\bf Remark 8.} It may be noted that a common time-dependent factor in the r.h.s. of \eqref{eq:Riccati} can be eliminated by a reparametrization of time; thus transformations multiplying the coefficients $f,g,h$ by a common factor should be seen as trivial, and one can restrict to consider $P : R \to SL(2,R)$. The connection between Riccati equations and gauge $SL(2,R)$ transformations has been widely studied, see e.g. \cite{RG1,RG2,RG3,RG4}. As far as we know, however, these studies did not put such a use of gauge transformations into a general context for generic ODEs. \EOR
\bigskip

We know that a gauge transform will map a linear system into a linear system, and that any linear system can be brought to autonomous form by a gauge transformation (see Remark 3 above); and we are of course able to solve a constant coefficients linear system. But the problem lies in determining the appropriate gauge transformation, i.e. in solving \eqref{eq:Pd} for the $A(t)$ at hand, and this could still turn out to be unfeasible. On the other hand, the equation can be solved in a number of cases.

\subsection{Example 8}

Let us consider the Riccati equation
\beql{eq:R11} \frac{dy}{dt} \ = \ \kappa_0 \, e^{- \b t} \ + \ \kappa_1 \, y \ + \ \( \kappa_2 \, e^{\b t} \) \, y^2 \ , \eeq
with $\kappa_i,\b$ real constants; we require $\kappa_2 \not= 0$, $\b \not= 0$ to discard trivial cases. We will also assume $\kappa_0 \not= 0$ (that is, we have a proper Riccati equation and not a Bernoulli equation). By a rescaling of time we can always set $\kappa_2 = 1$, which we do.

This Riccati equation corresponds to
\beq \dot{\wb} \ = \ A(t) \ \wb \ , \ \ A(t) \ = \ \begin{pmatrix} \kappa_1 & \kappa_0 \, e^{- \b t} \\
- e^{- \b t} & 0   \end{pmatrix} \ . \eeq
The gauge transformation with
\beq P(t) \ = \ \begin{pmatrix}  0 & \kappa_0 \\ - \, e^{\b t} & - \b - \kappa_1  \end{pmatrix} \eeq
(note that $\mathtt{Det} (P) = \kappa_0 e^{\b t}$ is nonzero by our assumption $\kappa_0 \not= 0$) maps this into
\beq B \ = \ \begin{pmatrix} \b + \kappa_1 & \kappa_0 \\ 1 & 0 \end{pmatrix} \ . \eeq This in turn corresponds to  the constants coefficients Riccati equation 
\beql{eq:R11T} \frac{dx}{dt} \ = \ r_0 \ + \ r_1 \, x \   -  \ x^2 \ , \eeq 
where the constants $r_1$ are given in terms of the constants $\{ \kappa_0,\kappa_1;\b\}$ appearing in the original equation (recall we set $\kappa_2 = 1$) by
\beq r_0 \ = \ \kappa_0 \ , \ \ r_1 \ = \ \kappa_1 \ + \ \b \ . \eeq

\subsection{Example 9}

Let us consider the Riccati equation
\beq \frac{dy}{dt} \ = \ 2 \, \sec (t) \ - \ \tan (t) \, y \ - \ \cos (t) \, y^2 \ . \eeq
This corresponds to
\beq \dot{\wb} \ = \ A(t) \ \wb \ , \ \ A(t) \ = \ \left(
\begin{array}{ll}
 -\tan (t) & 2 \sec (t) \\
 \cos (t) & 0
\end{array}
\right) \ .  \eeq

The gauge transformation with
\beq P \ = \ \left(
\begin{array}{ll}
 \frac{\tan (t)}{2} & -\sec (t) \\
 \frac{1}{2} & 0
\end{array}
\right) \eeq
maps this into the linear system with matrix
\beq B \ = \ \begin{pmatrix} 0 & -1 \\ -1 & 0 \end{pmatrix} \eeq
which corresponds to the constant coefficients Riccati equation
\beq \frac{dx}{dt} \ = \ - \, 1 \ + \ x^2 \ . \eeq
It should be noted that in this case
$$ \mathrm{Det} (P) \ = \ (1/2) \ \sec (t) \ , $$ so that the transformation is singular at $t = \pi/2 + k \pi$, and we can only work in time intervals of length $\pi$.

\subsection{Example 10}

Let us now consider the Riccati equation
\beq \frac{dy}{dt} \ = \ \( \frac{1-t}{1+t} \) \, \( \frac{2 + t}{t^2} \) \ + \ \( \frac{1}{1+t} \) \, \( \frac{2 - t^2}{t} \) \, y \ + \ \( 1 + t \) \, y^2   \ . \eeq
This is singular in $t=0$ and in $t = - 1$; we will hence think of this as being defined for $t > 0$. (We could of course also consider this for $t < -1$, or for $-1 < t < 0$.)

This equation corresponds to $\dot{\wb} = A \wb$ with
\beq A \ = \ \left(
\begin{array}{ll}
 \frac{2-t^2}{t (t + 1)} & \frac{2 - t - t^2}{t^2 (t+1)}
   \\
 -(t+1) & 0
\end{array}
\right) \ . \eeq
The gauge transformation with
\beq P \ = \ \left(
\begin{array}{ll}
 -\frac{t+1}{t} & -\frac{t+1}{t^2} \\
 1+\frac{1}{t} & \frac{1}{t^2}
\end{array}
\right) \ , \eeq
which is well defined in $t > 0$ (in fact, $\mathtt{Det} (P) = (1+t)/t^2$), takes this matrix into
\beq B \ = \ \begin{pmatrix} -1 & 1 \\ 1 & 0 \end{pmatrix} \ . \eeq
Thus we obtain in the end the constant coefficients Riccati equation
\beq \frac{dx}{dt} \ = \ 1 \ - \ x \ - \ x^2 \ . \eeq

\section{Discussion and conclusions}
\label{S:DC}

In this final Section we will first give a very brief discussion of (some) relations with the Physics literature and with PDEs \& Integrable Systems, in the form of Remarks, and then draw our conclusions.

\medskip\noindent{\bf Remark 9.} 
Our basic observation, as noted at the beginning of Sect.\ref{S:G}, is that the Floquet change of variables \eqref{eq:xtoy} can be given the interpretation of a gauge transformation. As remarked in Sect.\ref{S:G}, this seems not to have been discussed in the literature, at least in the framework of differential equations; things are different when we look at counterparts of Floquet's theory in Physics (Bloch theory).

In this context wide range investigations on the relation between periodicity (not only in time, but also in parameter space) and geometrical features have been carried out, also in relation to Berry phase and Hannay angle \cite{Berry,Hannay,Simon,Zak1,Zak2,SW,CF,ZKP,WZ,ASS,FCG,Mos}. It should be mentioned that these encompassed also non-periodic behavior \cite{Zak3,Zak4a,Zak4b,Pati}, in particular in the quasi-periodic case.\footnote{It is well known that discussing quasi-periodic linear operators and their spectra (in the Physics context one deals with second order operators, representing the Hamiltonian of the system under study) leads to difficult problems; see e.g.  \cite{Holst,DS,AAR,Eli,Sinai,Sim10}.}

See also, in a quite different physical context -- i.e. Fluid Mechanics -- the discussion in \cite{ArnKhe}. \EOR

\medskip\noindent
{\bf Remark 10.} 
Here we have been only concerned with dynamical systems; but it is
clear that the method of considering gauge transformations to
produce new systems amenable to some kind of exact treatment --
and conversely, reduce new type of systems to those for which an
exact treatment is available -- is more general and not limited
neither to dynamical systems nor to Floquet analysis; e.g. one
might produce new integrable systems out of known ones.

This remark can also be applied for PDEs, not just for Dynamical
Systems. E.g., if we consider the KdV equation \beql{eq:kdv} u_t \
= \ u_{xxx} \ + \ \kappa \, u \, u_x \eeq for $u = u(x,t)$ and
apply a gauge transformation \beql{eq:gkdv} u \ = \ \a (x,t) \, v
\ , \eeq with $\a$ a nowhere zero smooth function and $v = v(x,t)$
the new dependent variable, this is mapped into
\begin{eqnarray}
v_t &=& v_{xxx} \ + \ 3 \, \a^{-1} \, \( \a \, v_{xx} \ + \ \a_{xx} \, v_x \) \ + \ \a^{-1} \, \a_{xxx} \, v \nonumber \\
& & \ + \ \kappa \, \( \a_x \, v \ + \ \a v_x \) \, v \ - \
\a^{-1} \, \a_t \, v \ . \label{eq:kdv_t} \end{eqnarray} This is
by construction integrable, for \emph{any} smooth nowhere zero
function $\a (x,t)$, just inverting \eqref{eq:gkdv}; this fact is
not at all obvious when one looks at \eqref{eq:kdv_t} without
knowing how it was derived.

The interrelations between gauge transformations (or gauge
theories) and integrable systems -- in the sense of integrable
PDEs -- are of course well known and widely studied, see e.g.
\cite{Dun,DKN,DonWit,Ols}. It appears this was not considered in
the context of finite dimensional dynamical systems, considered
here. \EOR

\bigskip\noindent
{\bf Conclusions}

\medskip\noindent
We have considered the classical Floquet theory for periodic
linear systems of ODEs from a geometrical point of view, and
argued that the same approach can be extended to (linear) systems
with an arbitrary time dependence. The key concept is that of a
\emph{gauge transformation}; this corresponds to introducing a
nontrivial connection in the full phase space $R \times R^n$ seen
as a (trivial) bundle with base $R$ and fiber the reduced phase
space $R^n$. The main difference with standard Floquet theory is
that in that case, due indeed to periodicity, one can work in a
bundle over $S^1$ and is naturally led to consider a monodromy.

In this case, determination of the fundamental matrix in concrete
cases can be nontrivial; our approach allows, at least in
principles, to subdivide the difficulty of solving a
non-autonomous linear system in two steps, i.e. separate the
problem of determining the trajectories of the system (in
geometrical terms, covariantly constant sections of a bundle over
$R$) from that of the full dynamics, describing the time law on
these trajectories.

We have given several Examples, involving both Abelian and
non-Abelian gauge groups (i.e. gauge groups modelled on SO(2) and
SU(2) respectively), of applications of our approach to concrete
simple systems. In these Examples the equation \eqref{eq:Pd},
determining the appropriate gauge transformation converting the
time-dependent system into an autonomous one, could be determined
by inspection, thanks to their simple nature.

In Section
\ref{S:RE}, we have shown a concrete case in which our method can
be employed without \emph{apriori} information on the gauge group
to map a non-autonomous system of equations into an autonomous
one; this was done considering (the two-dimensional first order
system form of) Riccati equations.

In a companion paper of mathematical character \cite{GW2024}, we address the problem of
solving the equation \eqref{eq:Pd} in general cases, i.e. in
systems  not displaying such a simple form.

\label{lastpage}

\addcontentsline{toc}{section}{\ \ \ \ References}

\begin{thebibliography}{99}

\bibitem{GW2024} G. Gaeta and S. Walcher, ``On gauge transforms of autonomous ordinary differential equations'', 
Preprint {\it arXiv:2506.07189} (2025)


\bibitem{MarSka1} A.P. Balachandran, G. Marmo, B.S. Skagerstam, and A. Stern, {\it Gauge Symmetries and Fibre Bundles. Applications to
Particle Dynamics (Lect. Notes Phys. 188)}, Springer 1983

\bibitem{ArnODE} V.I. Arnold, {\it Ordinary Differential Equations}, Springer 1992

\bibitem{JKH} J.K. Hale, {\it Ordinary Differential Equations}, Dover 2009

\bibitem{FV} F. Verhulst, {\it Nonlinear Differential Equations and Dynamical Systems}, Springer 1990

\bibitem{Mermin} N.W. Ashcroft and N.D. Mermin, {\it Solid State Physics}, Saunders 1976

\bibitem{Kittel} Ch. Kittel {\it Introduction to Solid State Physics} Wiley  1986

\bibitem{Nak} M. Nakahara, {\it Geometry, Topology, and Physics}, IoP 1990

\bibitem{NS} C. Nash and S. Sen, {\it Topology and Geometry for Physicists}, Academic Press 1983

\bibitem{Schutz} B. Schutz, {\it Geometrical Methods of Mathematical Physics}, Cambridge UP, 1980

\bibitem{CCL} S.S. Chern, W.H. Chen and K.S. Lam, {\it Lectures on Differential Geometry}, World Scientific 1999

\bibitem{EGH} T. Eguchi, P.B. Gilkey and A.J. Hanson, ``Gravitation, gauge theories and differential geometry'', {\it Phys. Rep.} {\bf 66} (1980),  213-393



\bibitem{Berry} M.V. Berry, ``Quantal phase factors accompanying adiabatic changes'', {\it Proc. R. Soc. London A} {\bf 392} (1984), 45-57

\bibitem{Hannay} J.H. Hannay, ``Angle variable holonomy in adiabatic excursion of an integrable Hamiltonian'', {\it J. Phys. A} {\bf 18} (1985), 221-230

\bibitem{Simon} B. Simon, ``Holonomy, the quantum adiabatic theorem, and Berry's phase'', {\it Phys. Rev. Lett.} {\bf 51} (1983), 2167-2170

\bibitem{SW} A. Shapere and F. Wilczek eds., {\it Geometric Phases in Physics}, World Scientific 1989


\bibitem{Zak1} J. Zak, ``Berry's phase for energy bands in solids'', {\it Phys. Rev. Lett.} {\bf 62} (1989), 2747-2750

\bibitem{Zak2} J. Zak, ``Finite translations in solid-state physics'', {\it Phys. Rev. Lett.} {\bf 19} (1967), 1385-1387

\bibitem{CF} J. Cayssol and J.N. Fuchs, ``Topological and geometrical aspects of band theory'', {\it J. Phys. Materials} {\bf 4} (2021): 034007

\bibitem{ZKP} J.W. Zwanziger, M. Koenig, and A. Pines, ``Berry's phase'', {\it Ann. Rev. Phys. Chem.} {\bf 41} (1990), 601-646

\bibitem{WZ}   F. Wilczek and A. Zee, ``Appearance of gauge structure in simple dynamical systems'', {\it Phys. Rev. Lett.} {\bf 52} (1984), 2111-2114

\bibitem{ASS}  J.E. Avron, R. Seiler, and B. Simon, ``Homotopy and quantization in condensed matter physics'',  {\it Phys. Rev. Lett.} {\bf 51} (1983), 51-53

\bibitem{FCG} M. Fruchart, D. Carpentier and K. Gawedzki, ``Parallel transport and band theory in crystals'', {\it EuroPhys. Lett.} {\bf 106} (2014), 60002

\bibitem{Mos} A. Mostafazadeh, ``Non-Abelian geometric phase, Floquet theory and periodic dynamical invariants'', {\it J. Phys. A} {\bf 31} (1988), 9975-9982

\bibitem{Zak3} J. Zak, ``Berry's geometrical phase for noncyclic hamiltonians'', {\it Europhys. Lett.} {\bf 9} (1989), 615-620

\bibitem{Zak4a} J. Zak, ``Magnetic translation group'', {\it Phys. Rev.} {\bf 134} (1964), A1602-A1606

\bibitem{Zak4b} J. Zak, ``Magnetic translation group. II. Irreducible representations'', {\it Phys. Rev.} {\bf 134} (1964), A1607-A1611

\bibitem{Pati} A.K. Pati, ``Adiabatic Berry phase and Hannay angle for open paths'', {\it Ann. Phys.} {\bf 270} (1998), 178-197

\bibitem{WNM1} J. Wei and E. Norman, ``On global representations of the solutions of linear differential equations as a product of exponentials'', {\it Proc. A.M.S.} {\bf 15} (1964), 327-334

\bibitem{WNM2} J. Wei and E. Norman, ``Lie algebraic solution of linear differential equations'', {\it J. Math. Phys.} {\bf 4} (1963), 575-581

\bibitem{WNM3} J.F. Carinena, G. Marmo and J. Nasarre, ``The nonlinear superposition principle and the Wei-Norman method'', {\it Int. J. Mod. Phys. A} {\bf 13} (1998), 3601-3627

\bibitem{Ham} M. Hamermesh, {\it Group theory and its application to physical problems}, Addison-Wesley 1962

\bibitem{Gil} R. Gilmore, {\it Lie groups, Lie algebras, and some of their applications}, Wiley 1974

\bibitem{Kirillov} A.A. Kirillov, {\it Elements of the Theory of Representations}, Springer 2012

\bibitem{GRHH} G. Gaeta and M.A. Rodriguez, ``On the physical applications of hyper-Hamiltonian dynamics'', {\it  J. Phys. A} {\bf 41} (2008), 175203

\bibitem{Reid} W.T. Reid, ``A matrix differential equation of Riccati type'',
{\it Amer. J. Math.} {\bf 68} (1946), 237-246

\bibitem{RG1}  V M Strelchenya, ``A new case of integrability of the general Riccati
equation and its application to relaxation problems'', {\it J. Phys. A} {\bf 24} (1991), 4965-4967

\bibitem{RG2} J.F. Carinena and A. Ramos, ``Integrability of the Riccati equation from a group-theoretical viewpoint'', {\it Int. J. Mod. Phys. A} {\bf 14} (1999), 1935-1951

\bibitem{RG3} J.F. Carinena, J. De Lucas and A. Ramos, ``A geometric approach to integrability conditions for Riccati equations'', {\it Electronic J. Diff. Eqs.}, Vol. 2007 (2007), No. 122, pp. 1-14 (arXiv:0810.1740)

\bibitem{RG4} J.F. Carinena, J. Grabowski, and G. Marmo, ``Superposition rules, Lie theorem, and partial differential equations'', {\it Rep. Math. Phys.} {\bf 60} (2007), 237-258

\bibitem{ArnKhe} V.I. Arnold and B. Khesin, {\it Topological methods in hydrodynamics}, Springer 2009

\bibitem{Holst} T. Holstein, ``Studies of polaron motion, part 1. The molecular-crystal model'', {\it Ann. Phys.} {\bf  8} (1959), 325-342

\bibitem{DS} E.I. Dinaburg and Ya.G. Sinai, ``On the one dimensional Schroedinger equation with a quasiperiodic potential'', {\it Funct. Anal. and its Appl.} {\bf 9} (1975), 279-289

\bibitem{Sinai} Ya.G. Sinai, ``Anderson localization for one-dimensional difference Schroedinger operator with
quasiperiodic potential'', {\it J. Stat. Phys.} {\bf 46} (1987), 861-909

\bibitem{Eli} L.H. Eliasson, ``Floquet Solutions for the 1-Dimensional Quasi-Periodic Schroedinger Equation'', {\it Commun. Math. Phys.} {\bf 146} (1992), 447-482

\bibitem{AAR} S. Aubry, G. Abramovici, and J. Raimbaut, ``Chaotic polaronic and bipolaronic states in the adiabatic Holstein model'', {\it J. Stat. Phys.} {\bf 67} (1992), 675-780

\bibitem{Sim10} B. Simon, ``Almost periodic Schroedinger operators: A review'', {\it Adv. Appl. Math.} {\bf 3} (1982), 463-490

\bibitem{Dun} M. Dunajski, {\it Solitons, instantons, and twistors}, Oxford UP 2024

\bibitem{DKN} B.A. Dubrovin, I.M. Krichever, and S.P. Novikov, {\it Integrable systems. I}  (Dynamical Systems IV), Springer  1990

\bibitem{DonWit} R. Donagi and E. Witten, ``Supersymmetric Yang-Mills theory and integrable systems'', {\it Nucl. Phys. B} {\bf 460} (1996), 299-334

\bibitem{Ols} M. Olshanetsky, ``Classical integrable systems and gauge field theories'', {\it Phys Part. Nuclei} {\bf 40} (2009), 93-114



\end{thebibliography}
\end{document}